\documentclass[journal]{IEEEtran}
\usepackage{algorithm}
\usepackage{algpseudocode}
\usepackage{graphicx}
\usepackage{hyperref}
\usepackage{verbatim} 
\usepackage{multirow} 
\usepackage{amsmath}
\usepackage{multirow}

\usepackage{booktabs} 

\ifCLASSINFOpdf
\else
\fi
\begin{document}
%
\title{Accelerating 5G Synchronization Signal Timing Offset Estimation Using Dual-Rate Sampling}

%
%
%

\author{Bitna~Kim,~\IEEEmembership{Member,~IEEE,}
        Seungyeon~Lee, ~\IEEEmembership{Student member,~IEEE,}
        Yelan~Lee, ~\IEEEmembership{Student member,~IEEE,}
        and~Juyeop~Kim ~\IEEEmembership{Member,~IEEE,}
\thanks{B. Kim is with Hyundai Mobis, and S. Lee, Y. Lee and J. Kim are with the Department of Electrical Engineering at Sookmyung Women's University (e-mail: bn7952@naver.com, songee2002@sookmyung.ac.kr, yr00868@sookmyung.ac.kr and jykim@sookmyung.ac.kr). Juyeop Kim is the corresponding author.}
}

\maketitle

\begin{abstract}
Cell search engineers face significant challenge in reducing computation time to meet the requirements for fast initial access and radio link recovery.
Since the majority of cell search time is consumed by Primary Synchronization Signal (PSS) detection, reducing the computational burden of this step is critical for shortening the overall procedure.
This paper proposes a novel timing offset estimation scheme designed to accelerate 5G cell search.
Leveraging the 5G Synchronization Signal Block (SSB) structure, the proposed scheme employs a two-step estimation process using dual-rate sampling. This approach effectively reduces the PSS detection search space without compromising the performance of subsequent processes.
Performance evaluations in practical system and channel environments demonstrate that the proposed scheme reduces the cell search procedure time by 68\% compared to the baseline, while maintaining Physical Broadcast CHannel (PBCH) decoding performance.
\end{abstract}

\begin{IEEEkeywords}
5G, initial cell search, SSB, primary synchronization signal, timing offset estimation.
\end{IEEEkeywords}

%
\IEEEpeerreviewmaketitle

\section{Introduction}

\IEEEPARstart{C}{ell} search, which aims to identify cells and their synchronization parameters, has been a critical component across all generations of mobile communications.
Due to the stringent performance requirements for accuracy and time efficiency, signal detection techniques within the cell search procedure have become increasingly sophisticated \cite{cellSearch}.
Consequently, many systems, including 5G New Radio (NR), employ dual Synchronization Signals (SSs) --- the Primary Synchronization Signal (PSS) and the Secondary Synchronization Signal (SSS) --- to enable multi-step searching across independent dimensions \cite{LTEcellSearch, 5GcellSearch}.
These structured signals allow User Equipments (UEs) to decouple time-consuming timing offset estimation from the broader cell search procedure, thereby mitigating its impact on subsequent reception performance.

Conventional studies have contributed to enhancing cell search by utilizing dual SSs \cite{conv_cs_total}.
Building upon the traditional correlation-based approaches for Orthogonal Frequency Division Multiplexing (OFDM) systems \cite{legacy1, legacy2}, subsequent research has sought to achieve more accurate timing offset estimation.
For instance, the specific structure of repeated SSs was designed to facilitate stochastic estimation while effectively suppressing noise \cite{PSS_2}.
Furthermore, integer Carrier Frequency Offset (CFO) is typically addressed prior to PSS detection by exploiting the cyclic characteristics of PSS sequences \cite{PSS_1}.
This CFO is then further compensated using the PSS-derived estimate, thereby reducing phase distortion and enabling more accurate SSS detection \cite{conv_CFO2}.


Extending these conventional approaches, recent studies have addressed synchronization challenges within the 5G and 6G ecosystems.
Specifically, more accurate estimation of both integer and fractional CFO has been proposed for non-terrestrial networks by leveraging the PSS \cite{conv_CFO_LEO, conv_CFO_LEO2}.
Additionally, new synchronization schemes have been developed to mitigate interference in cell-free massive MIMO environments \cite{conv_Sync_6G}.
To bridge the gap between theory and practice, low-complexity detection procedures for dual SSs have been softwarized and verified within commercial 5G infrastructure \cite{sync_6G_3}.
Furthermore, the performance of SS detection has been significantly enhanced through Artificial Intelligence (AI), with its feasibility demonstrated particularly in low-SNR regimes \cite{conv_AI1, conv_AI2, conv_AI3}.

While the SSs are designed for efficient timing offset estimation, their detection process has recently become increasingly computationally intensive.
Due to the wider bandwidths involved, 5G UEs operate at higher sampling rates, which significantly expands the search space for timing offset estimation.
Although the increased bandwidth of the SSs is intended to improve auto-correlation characteristics, it inherently raises the computational complexity of the estimation process.
Consequently, managing this complexity has emerged as a critical challenge for next-generation systems.
Several studies have attempted to reduce the burden of correlation-based timing offset estimation; however, these approaches often suffer from degraded detection probability or prove impractical for real-time cell search scenarios \cite{PSS_LowComp1, PSS_LowComp2, PSS_LowComp3}.

To address these challenges in 5G systems, this paper proposes a timing offset estimation scheme based on the structure of the Synchronization Signal Block (SSB).
The proposed scheme employs dual sampling rates --- specifically, half and full rates --- each strategically selected for the corresponding SS within the SSB to minimize overall detection complexity.
The following sections present the signal model of the 5G SSB and describe how the proposed scheme achieves low-complexity timing offset estimation.
Furthermore, we demonstrate how the proposed approach contributes to a reduction in detection time through performance evaluations conducted in over-the-air signal environments.
\section{System Model for SSB Detection}

\begin{figure}[t]
\centerline{\includegraphics[width=0.9\linewidth]{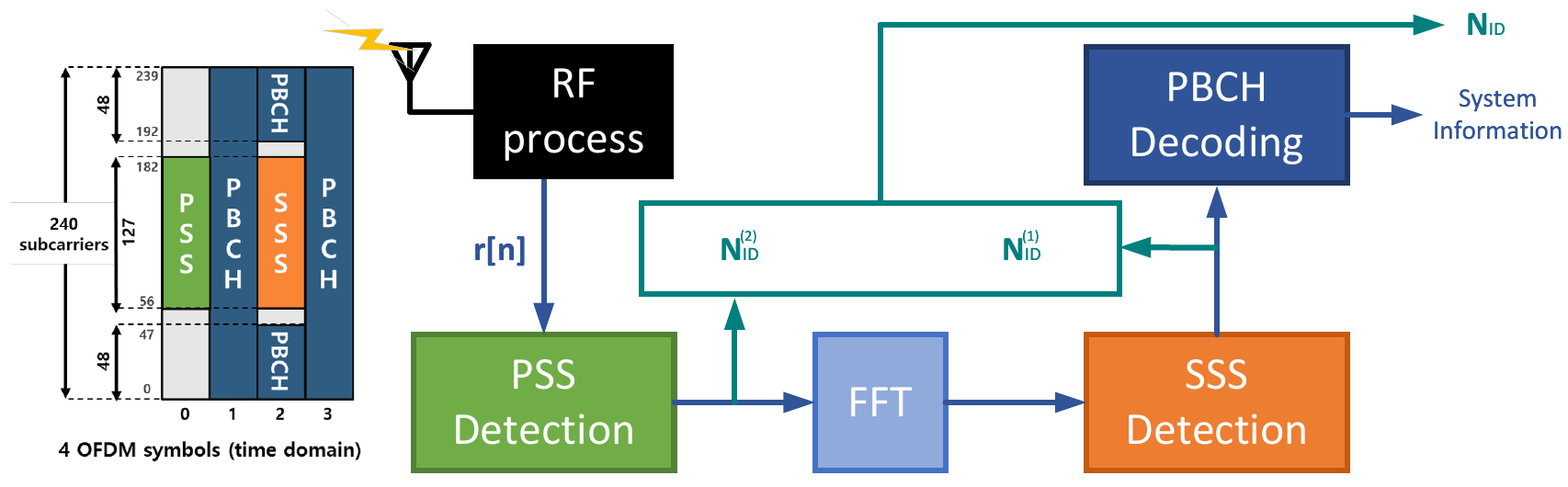}}
\caption{The SSB Structure and the 5G Cell Search Procedure} 
\label{Fig:CellSearch}
\end{figure}

We assume the frame parameters and the associated SS structure defined in the 5G NR standard.
The sampling rate, denoted by $f_s$, is given by the product of the Fast Fourier Transform (FFT) size $N_{\text{FFT}}$ and the subcarrier spacing $\Delta f$.
Based on these parameters, the 5G gNodeB (gNB) with a cell ID $N_{\text{ID}}$ transmits an SSB with a period of $T_{\text{SSB}}$. The structure of the SSB is illustrated in Fig.~\ref{Fig:CellSearch}.
The SSB comprises the Physical Broadcast CHannel (PBCH), which carries system information, along with the PSS and SSS transmitted over four contiguous OFDM symbols.
Specifically, the PSS and SSS are mapped to the central half of the SSB bandwidth, whereas the PBCH spans the entire SSB bandwidth.
The 5G NR standards specifies three distinct PSS sequences; we denote by $s_i[n]$ the transmitted time-domain samples of the first OFDM symbol carrying the $i$-th PSS sequence.
Fig.~\ref{Fig:CellSearch} illustrates the overall cell search procedure considered in this paper \cite{sync_6G_3}.
At a given center frequency, the RF processing block converts the received RF signal from the antenna into a baseband signal sampled at rate of $f_s$.
Based on these baseband samples, the primary synchronization step estimates the timing offset of the SSB by detecting the PSS.
Subsequent steps process the SSS and PBCH using frequency-domain symbols derived from the estimated timing offset.
The PSS and SSS detection results are then used to determine $N_{\text{ID}}$, while PBCH decoding provides the system information of the gNB.

After the UE obtains the received signal samples, denoted by $r[n]$, it first detects the PSS component within the received SSB by computing the cross-correlation with $s_i[n]$.
Through this PSS detection, the UE estimates the timing offset $\tau_{\text{SSB}}$ and the PSS sequence index $N_{\text{ID}}^{\text{(2)}}$ as follows:
\begin{eqnarray}
(\tau_{\text{SSB}},N_{\text{ID}}^{\text{(2)}}) &=& \arg\max_{\tau \in {\textbf{N}_{\text{PSS}}},i} \rho_i[\tau], \label{acorr1}\\
\rho_i[\tau] &=& \left|\sum_{k=1}^{N_{\text{FFT}}} r[\tau-1+k] \cdot {s_{i}[k]} \right|,\label{acorr2}
\end{eqnarray}
where $\textbf{N}_{\text{PSS}}=\{1,2,\cdots,N_{\text{SSB}}\}$ denotes the set of candidate timing offsets within the SSB periodicity, where $N_{\text{SSB}} = T_{\text{SSB}}f_s$.
Subsequent steps process SSS and PBCH using frequency-domain symbols extracted based on the estimated $\tau_{\text{SSB}}$.
For these subsequent procedures, we adopt the specific algorithms provided by OpenAirInterface (OAI) \cite{sync_6G_3}.

Timing offset estimation in (\ref{acorr1}) and (\ref{acorr2}) accounts for the major portion of the overall cell search time, especially when the search space $\textbf{N}_{\text{PSS}}$ is large.
From a computational perspective, the cost of acquiring a single correlation value $\rho_i[\tau]$ is $\mathcal{O}\left(N_{\text{FFT}}\right)$.
Since the estimation in (\ref{acorr1}) requires computing $\rho_i[\tau]$ for all $i\in\{0, 1, 2\}$ and all candidate offsets $\tau \in \textbf{N}_{\text{PSS}}$, the total computational complexity, denoted by $Q_{s}(f_s)$, can be expressed as a function of $f_s$ as follows:
\begin{eqnarray}
Q_{s}(f_s) &=& \mathcal{O}\left(3N_{\text{SSB}}f_sN_{\text{FFT}}^2\right) = \mathcal{O}\left(3N_{\text{SSB}}\frac{f^2_s}{\Delta f}\right).\label{cmp_conv}
\end{eqnarray}
(\ref{cmp_conv}) demonstrates that the complexity of timing offset estimation is proportional to the size of the search space and, more critically, to the square of the sampling rate.
\section{Two-Step Timing Offset Estimation with Dual Sampling Rates}

An effective way to reduce overall cell search time is to minimize the computational complexity of timing offset estimation.
As shown in (\ref{cmp_conv}), the computations in (\ref{acorr1}) can be significantly simplified by reducing the sampling rate.
However, a lower sampling rate curtails the time resolution of $r[n]$, which may degrade detection performance in subsequent cell search steps.
Therefore, careful selection of the sampling rate is essential to balance low computational complexity against reliable detection performance.

Based on the above observation, the proposed scheme achieves low computational complexity by sampling the received signal at two different rates.
As illustrated in Fig.~\ref{Fig:CellSearch}, the PSS occupies only half of the total bandwidth allocated to the SSB.
This indicates that PSS detection requires only half the sampling rate necessary for PBCH decoding.
Consequently, sampling the entire SSB at a unique rate would lead to an unnecessary increase in computational complexity during the PSS detection step.
To minimize this overhead, it is preferable to employ decoupled sampling rates tailored to the specific requirements of PSS detection and PBCH decoding.

\begin{figure}[t]
\centerline{\includegraphics[width=1\linewidth]{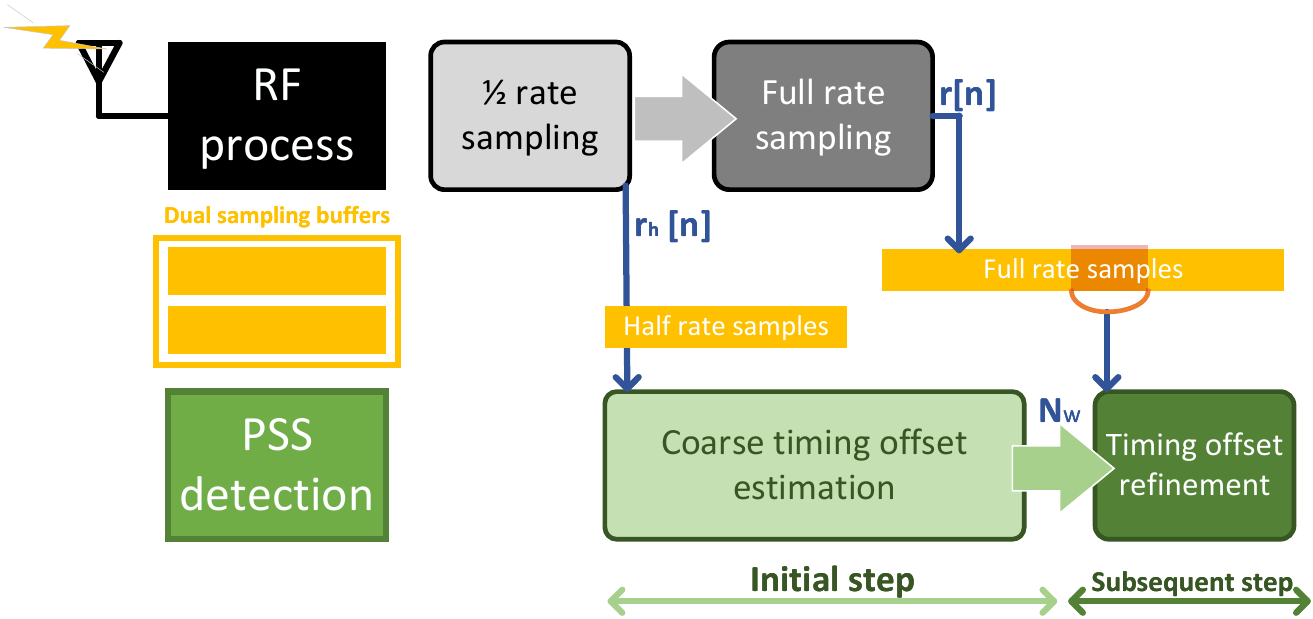}}
\caption{Two-step Timing Offset Estimation Process} 
\label{Fig:propsedscheme} 
\end{figure}

Fig.~\ref{Fig:propsedscheme} illustrates how the proposed scheme estimates the timing offset using dual sampling rates.
The novelty lies in the two estimation sub-steps, each of which utilizes a different sampling rate optimized for its specific function.
In the initial sub-step, the scheme estimates a coarse timing offset using half-rate sampling.
Simultaneously, the proposed scheme performs full-rate sampling to provide the high-resolution data required for the next sub-step.\footnote{The system in Fig.~\ref{Fig:propsedscheme} is assumed to perform timing offset estimation and sampling in parallel at the baseband and RF processors, respectively.}
Let $r_h[n]$ denote the half-rate samples, where $n \in \textbf{N}_h$ and $\textbf{N}_h = \{1, \cdots, N_{\text{SSB}}/2\}$.
Based on the coarse timing offset, the proposed scheme then constructs a localized search range and estimates the precise timing offset within this range using the full-rate samples.

In the initial sub-step, the coarse timing offset $\tau_{h}$ is estimated from the half-rate samples as follows:
\begin{eqnarray}
(\tau_{h},N_{\text{ID}}^{\text{(2)}}) &=& \arg\max_{\tau \in {\textbf{N}_h},i} \rho_{h,i}[\tau], \label{acorr1_prop}\\
\rho_{h,i}[\tau] &=& \left|\sum_{k=1}^{N_{\text{FFT}}/2} r_h[\tau-1+k] \cdot {s_{h,i}[k]} \right|,\label{acorr2_prop}
\end{eqnarray}
where $s_{h,i}[n]$ denotes the transmitted half-rate samples of the first OFDM symbol carrying the $i$-th PSS sequence.
The search space for (\ref{acorr1_prop}) is only one-fourth that of (\ref{acorr1}).
This reduction occurs because both the search space $\textbf{N}_h$ and the reference sequence $s_{h,i}[n]$ are half the length of their full-rate counterparts, $\textbf{N}_{\text{PSS}}$ and $s_{i}[n]$.
This complexity reduction is consistent with (\ref{cmp_conv}), which demonstrates that the computational load scales with the square of the sampling rate.

Once $\tau_{h}$ is obtained, the subsequent sub-step is performed immediately as the acquision latency of the initial sub-step significantly exceeds the time required to collect the full-rate samples.
To account for possible errors in the coarse timing offset --- arising from estimation inaccuracies or clock drift --- the proposed scheme performs a refined search within a margin $\Delta n$ as follows:
\begin{eqnarray}
\tau_{\text{SSB}} &=& \arg\max_{\tau \in {\textbf{N}_{w}}} \rho_{N_{\text{ID}}^{\text{(2)}}}[\tau], \label{acorr_normal}\\
\textbf{N}_w &=& \{\tau_h-\Delta n, \cdots \tau_h+\Delta n\}. \label{accor_range}
\end{eqnarray}
The margin $\Delta n$ is determined by a rule of thumb based on the coherent time of the target channels and the rate-conversion characteristics of the RF hardware.
This sub-step incurs negligible computational overhead compared to the initial sub-step, provided that $\Delta n$ is sufficiently smaller than $N_{\text{SSB}}$ and the search space in (\ref{acorr_normal}) remains constrained.

\begin{algorithm}[t] 
\caption{Two-Step Timing Offset Estimation} 
\label{algorithm} 
\small 
\begin{algorithmic}[1] 

\State \textbf{1. Half-rate sampling}: 
\State Sample the received signal with the half rate $f_s/2$ in $r_h[n]$.

\State \textbf{2. Initial-step estimation}: 
\State Sample the received signal with the full rate $f_s$ in $r[n]$.
\State Estimate coarse timing offset $\tau_h$ as follows:
\State $\rho_{\text{max}} = 0$.
\For{$\tau = 1$ to $N_{\text{SSB}}/2$}
    \For{$i = 0$ to $2$}
        \State $\rho_{h,i}[\tau] = 0$
        \For{$k = 1$ to $N_{\text{FFT}}/2$}
            \State $\rho_{h,i}[\tau] = \rho_{h,i}[\tau] + r_h[\tau - 1 + k] s_{h,i}[k]$ 
        \EndFor
        
        \If{$|\rho_{h,i}[\tau]| > \rho_{\text{max}}$} 
            \State $\rho_{\text{max}} = |\rho_{h,i}[\tau]|$
            \State $\tau_h = \tau$.
            \State $N_{\text{ID}}^{\text{(2)}} = i$. 
        \EndIf 
    \EndFor 
\EndFor 

\State \textbf{3. Subsequent-step refinement}:
\State $\rho_{\text{max}} = 0$.
\For{$\tau = \tau_h - \Delta n$ to $\tau_h + \Delta n$}
    \State $\rho_{N_{\text{ID}}^{\text{(2)}}}[\tau] = 0$
    \For{$k = 1$ to $N_{\text{FFT}}/2$}
        \State $\rho_{N_{\text{ID}}^{\text{(2)}}}[\tau] = \rho_{N_{\text{ID}}^{\text{(2)}}}[\tau] + r[\tau - 1 + k] s_{i}[k]$ 
    \EndFor
    
    \If{$|\rho_{N_{\text{ID}}^{\text{(2)}}}[\tau]| > \rho_{\text{max}}$} 
        \State $\rho_{\text{max}} = |\rho_{N_{\text{ID}}^{\text{(2)}}}[\tau]|$
        \State $\tau_{\text{SSB}} = \tau$.
    \EndIf 
\EndFor 

\end{algorithmic} 
\end{algorithm}

Algorithm.~\ref{algorithm} illustrates the overall procedure of the proposed timing offset estimation.
Following up half-rate sampling, the initial sub-step performs full-rate sampling and coarse timing offset estimation in parallel.
Assuming that full-rate sampling requires less time than the coarse estimation process, the computational complexity of this sub-step is $\mathcal{O}\left(\frac{3N_{\text{SSB}}}{4}\frac{f_s}{\Delta f}\right)$.
The subsequent sub-step refines the timing offset within a range of length $2\Delta n$, with a computational complexity of $\mathcal{O}\left(2\Delta n\frac{f_s}{\Delta f}\right)$.
If $\Delta n$ is configured to be significantly smaller than $N_{\text{SSB}}$ and $\Delta n << \frac{3}{8}N_{\text{SSB}}$, the complexity of the refinement sub-step becomes negligible.
In this case, the total computational complexity of the proposed scheme is $\mathcal{O}\left(\frac{3N_{\text{SSB}}}{4}\frac{f_s}{\Delta f}\right)$, which represents 75\% reduction (or is one-fourth) compared to the baseline.

The estimation performance of the proposed scheme depends on the time-varying characteristics of the channel.
The timing offset refinement in the subsequent sub-step remains valid only if the true timing offset falls within the range $\textbf{N}_w$.
In environments with severe channel fading or high mobility, the UE may experience a significant time shift between the initial and subsequent sub-steps; if this shift exceeds the search margin, the UE will fail to detect the PSS within $\textbf{N}_w$.
Consequently, $\Delta n$ must be configured with a sufficiently large value to prevent detection failure caused by time-varying channel conditions.
This selection is not critical to the overall computational complexity, as the time interval between the two estimation sub-steps is typically $T_{\text{SSB}} = 20\text{ms}$ in 5G.\footnote{Commercial UEs are capable of tracking synchronization within this interval across various practical channel environments.}
\begin{figure*}[t]
\centerline{\includegraphics[width=0.80\linewidth]{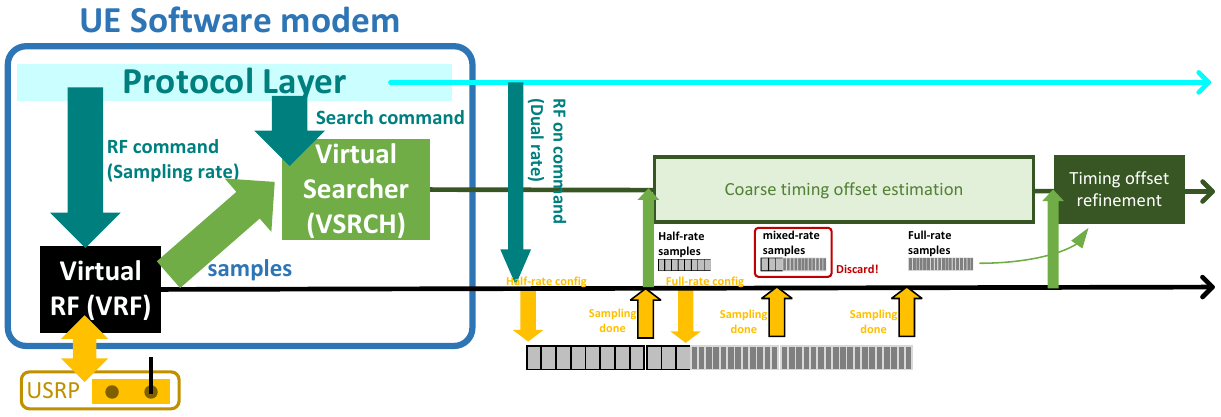}}
\caption{The Mechanism of the proposed scheme in the UE Software Modem}
\label{Fig:SWModem} 
\end{figure*}
\begin{figure}[t]
\centerline{\includegraphics[width=0.95\linewidth]{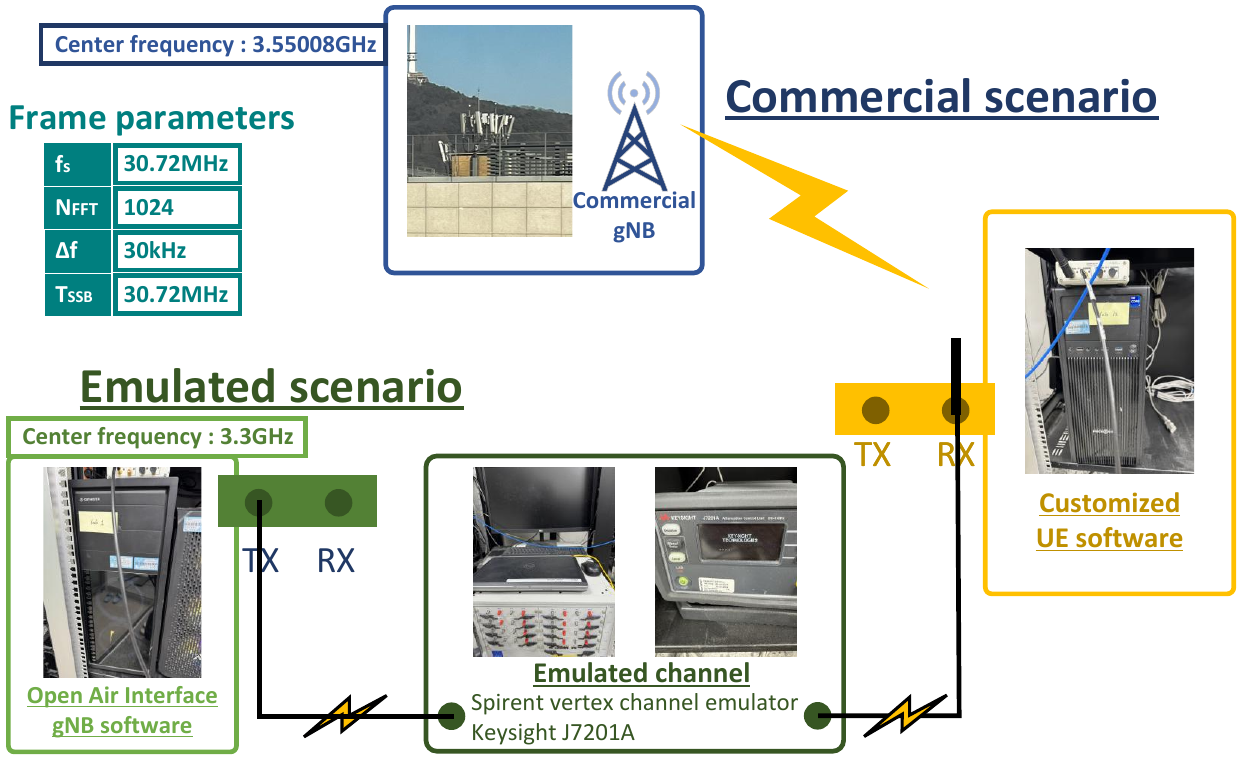}}
\caption{The Testbed System with Two Test Scenarios} 
\label{Fig:TBsystem} 
\end{figure}


\section{Testbed Implementation and Performance Evaluation}

For practical evaluation of time consumption and detection performance, we utilized an off-the-shelf testbed to conduct 5G cell search procedures in real-time.
Fig.~\ref{Fig:TBsystem} illustrates the testbed system, which incorporates commercial and emulated gNBs, a UE, and their frame parameters.
The emulated gNB and UE each consist of a USRP B210 and a Linux PC running a software modem on an Intel Core i7-8700K CPU with a clock speed of 3.70GHz.
The UE repeats 5G cell search procedures in the following two test scenarios: 
\begin{itemize}
\item Emulated scenario : Cell search via the emulated gNB with varying channel environments provided by channel emulation devices.
\item Commercial scenario : Cell search via a commercial SK Telecom gNB at various locations with different channel environments.
\end{itemize}

The emulated gNB runs the OAI software and generates baseband signals based on its native algorithms \cite{OAI}.
The UE runs a customized software modem incorporating the proposed scheme \cite{sync_6G_3}.
Fig.~\ref{Fig:SWModem} illustrates the detailed mechanism of this scheme within the UE software modem.
The Virtual RF (VRF), which configures the USRP sampling rate, transitions from half-rate to full-rate immediately following the half-rate sampling to accelerate the refinement sub-step.
Because the VRF switches to the full-rate midway through the half-rate sampling process, it discards any mixed-rate samples and delivers the subsequent full-rate samples to the Virtual Searcher (VSRCH).
This pipelining ensures that full-rate samples are available earlier, allowing the VSRCH to begin the refinement sub-step immediately without waiting for additional buffering.

Table.~\ref{tab:cell_search_time} presents the average time consumption for the overall cell search procedure and each constituent step.
The overall cell search using the proposed scheme consumes only 31.7\% of the time required by the baseline.
The primary driver of this reduction is the disproportionate impact of PSS detection time on the total process; the baseline requires more than 99\% of its total time simply to detect the PSS.
This high latency occurs because PSS detection is the initial step of the procedure, generally requiring blind detection across a massive search space.
While the proposed scheme introduces a small amount of additional time for timing offset refinement, this sub-step features low computational complexity and has a negligible impact on the overall latency.
We can conclude that dual-rate sampling enables the proposed scheme to effectively detect the PSS within a more compact search space, thereby drastically reducing the overall cell search time.

\begin{figure}[h]
\centerline{\includegraphics[width=0.95\linewidth]{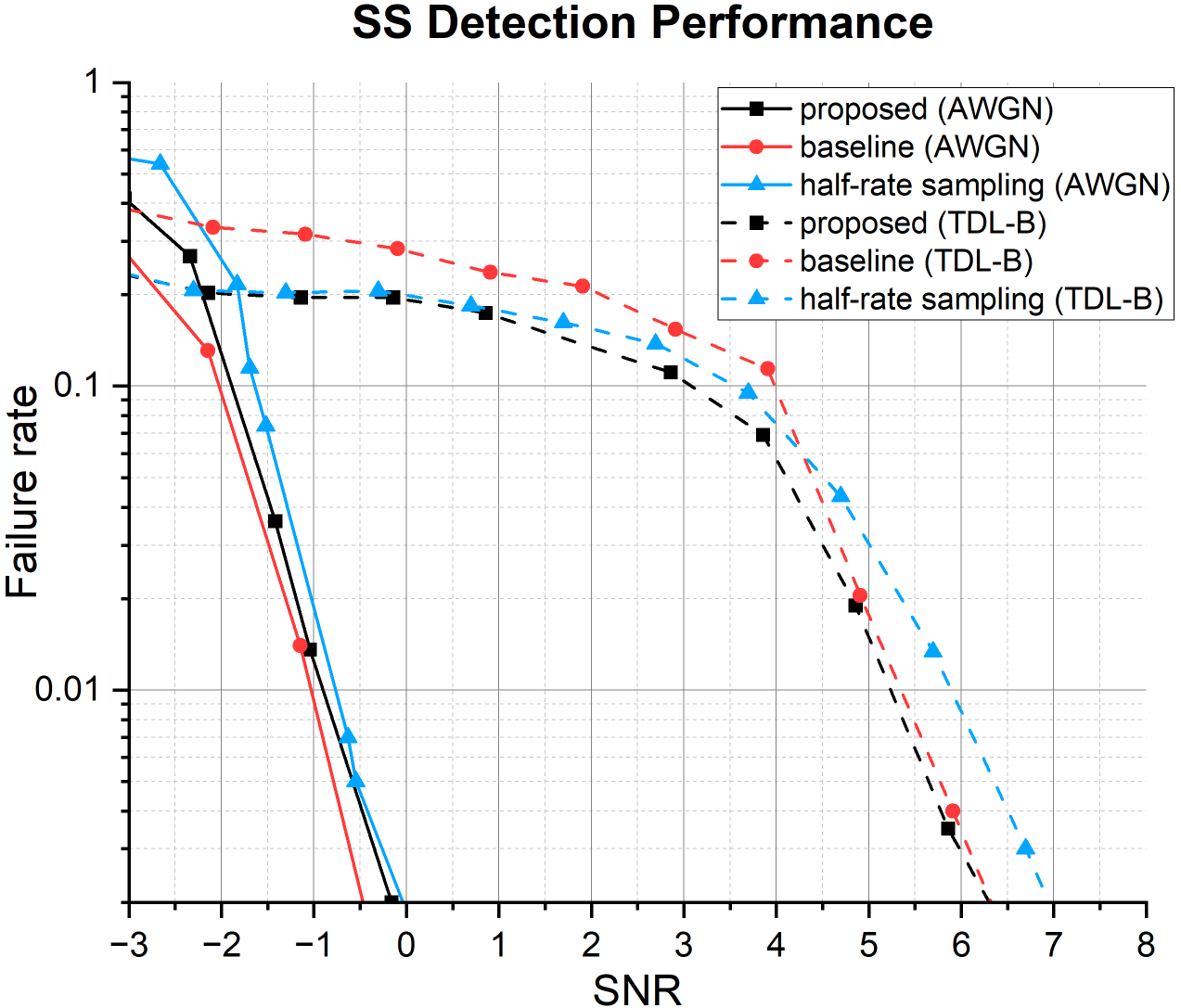}}
\caption{Performance Comparison from a Signal Detection Perspective}
\label{Fig:SSdet} 
\end{figure}

\begin{figure}[h]
\centerline{\includegraphics[width=0.95\linewidth]{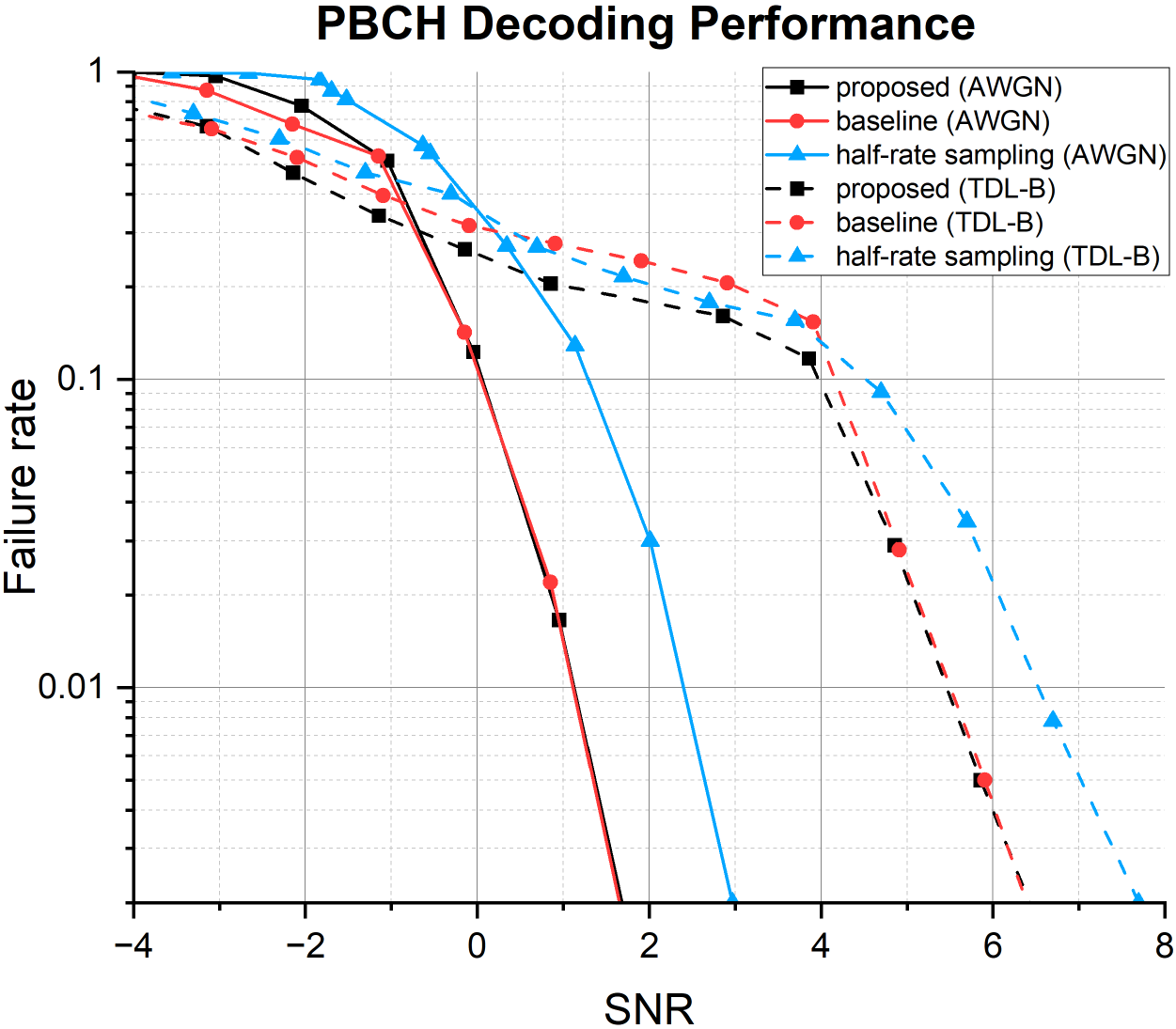}}
\caption{Performance Comparison from a Decoding Perspective}
\label{Fig:pbchdec} 
\end{figure}

Fig.~\ref{Fig:SSdet} and ~\ref{Fig:pbchdec} illustrate how the various schemes impact detection performance in AWGN and time-varying fading channels.
For the fading case, we implemented a time-varying environment by configuring the channel emulator to Tapped Delay Line (TDL)-B 30 km/h.
Fig.~\ref{Fig:SSdet} compares the probability of cell ID detection failure.
It reveals that the proposed scheme achieves similar cell ID detection performance to the other schemes in both AWGN and time-varying channels.
This confirms that the two-step PSS detection employed by the proposed scheme does not degrade detection accuracy, performing comparably to the baseline single-step PSS detection.
Furthermore, Fig.~\ref{Fig:pbchdec} compares the probability of PBCH decoding failure, proving that the proposed scheme yields SNR gains of 1 -- 1.5 dB over the half-rate sampling scheme.

\begin{figure}[h]
\centerline{\includegraphics[width=0.95\linewidth]{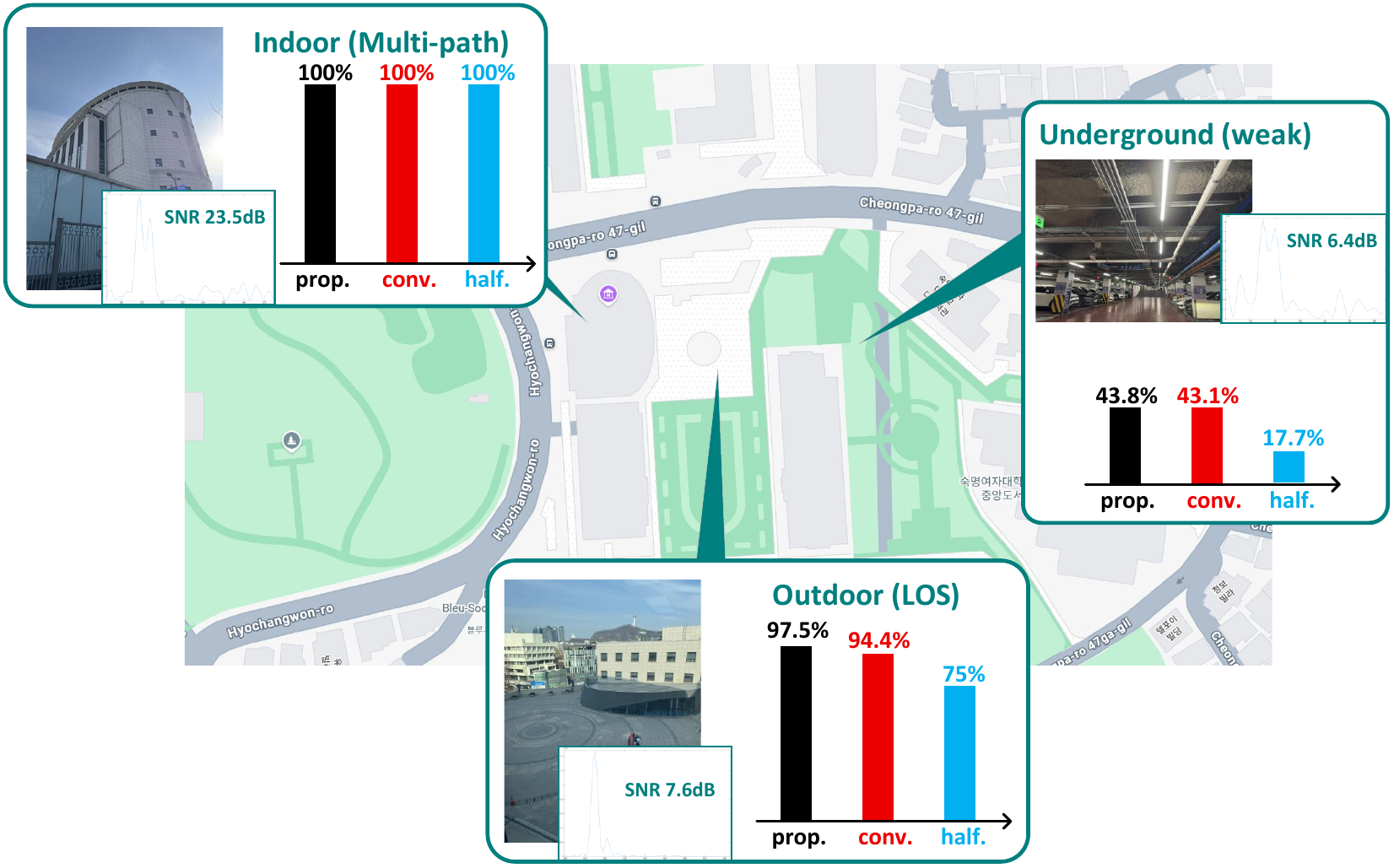}}
\caption{PBCH Decoding Performance in Field Environments}
\label{Fig:field} 
\end{figure}

Fig.~\ref{Fig:field} illustrates the PBCH decoding performance in various static field environments where the UE conducts cell search for commercial gNBs.
As observed in Fig.~\ref{Fig:SSdet} and ~\ref{Fig:pbchdec}, the proposed scheme achieves performance comparable to the full-rate sampling scheme, while outperforming the half-rate sampling scheme in terms of PBCH decoding.
On the other hand, the proposed scheme completes the cell search in significantly less time than the full-rate sampling scheme as shown in Table.~\ref{tab:cell_search_time}.
These results confirm that the proposed scheme efficiently performs cell search from a time-consumption perspective without any degradation in detection performance.

\begin{table}[t]
\caption{Average Time Consumption of the 5G Cell Search Procedure}
\label{tab:cell_search_time}
\centering
\begin{tabular}{|c|c|c|c|c|c|}
\hline
Time (ms) & \textbf{Overall} & \textbf{PSS (init)} & \textbf{PSS (refine)} & \textbf{SSS} & \textbf{PBCH} \\ 
\hline
\textbf{Baseline} 
& 142.75 
& 141.82 
& -- 
& \multirow{2}{*}{0.528} 
& \multirow{2}{*}{0.393} \\ 
\cline{1-4}
\textbf{Proposed} 
& 45.30  
& 38.26  
& 5.85 
&  
&  \\ 
\hline
\end{tabular}
\end{table}

\section{Conclusion}
Minimizing cell search time is critical in modern mobile communications as the search space continuously expands.
Based on the signal structure of the SSB, we propose a two-step timing offset estimation scheme with dual-rate sampling to accelerate the overall cell search procedure.
The proposed scheme is designed to efficiently acquire and process half-rate and full-rate samples in parallel, contributing to a significant reduction in cell search time.
Performance evaluation using a real-time testbed reveals that the proposed scheme efficiently detects SSBs in a shorter time without causing performance degradation in practical environments.


%

\bibliographystyle{IEEEtran}
\bibliography{6_ref}

\ifCLASSOPTIONcaptionsoff
  \newpage
\fi


\end{document}